\documentclass{article}
\usepackage{subcaption}

\usepackage[english]{babel}

\usepackage[letterpaper,top=2cm,bottom=2cm,left=3cm,right=3cm,marginparwidth=1.75cm]{geometry}

\usepackage{amsmath}
\usepackage{graphicx}
\usepackage[colorlinks=true, allcolors=blue]{hyperref}
\usepackage{comment}
\usepackage{authblk}
\usepackage{color}
\usepackage{booktabs}
\usepackage{orcidlink}
\usepackage[commandnameprefix=always]{changes}

\title{PWACG: Partial Wave Analysis Code Generator supporting Newton-conjugate gradient method}
\author[1]{Xiang Dong\thanks{These authors contributed equally to this work.}\orcidlink{0009-0004-3851-2674}}
\author[1]{Yu-Chang Sun\textsuperscript{*}\orcidlink{0009-0009-8756-8718}}%
\author[1]{Chu-Cheng Pan\orcidlink{0009-0009-9985-9950}}
\author[1]{Ao-Yan Cheng\orcidlink{0009-0009-2622-2721}}
\author[1]{Ao-Bo Wang\orcidlink{0009-0007-1036-1570}}
\author[1]{Hao Cai\thanks{Corresponding author: hcai@whu.edu.cn}\orcidlink{0000-0003-0898-3673}}
\author[2]{Kai Zhu\thanks{Corresponding author: zhuk@ihep.ac.cn}\orcidlink{0000-0002-4365-8043}}
\affil[1]{Wuhan University, Wuhan 430072, People's Republic of China}
\affil[2]{Institute of High Energy Physics, Beijing 100049, China}

\begin{document}
\maketitle

\begin{abstract}
This paper introduces a novel Partial Wave Analysis Code Generator (PWACG) that automatically generates high-performance partial wave analysis codes. This is achieved by leveraging the JAX automatic differentiation library and the jinja2 template engine. The resulting code is constructed using the high-performance API of JAX, and includes support for the Newton's Conjugate Gradient optimization method, as well as the full utilization of parallel computing capabilities offered by GPUs. By harnessing these advanced computing techniques, PWACG demonstrates a significant advantage in efficiently identifying global optimal points compared to conventional partial wave analysis software packages.
\end{abstract}


\section*{Program Summary}
\begin{itemize}
    \item \textbf{Program Title:} PWACG: Partial Wave Analysis Code Generator
    \item \textbf{Licensing provisions:} This software is distributed under the MIT License.
    \item \textbf{Programming language:} Python
    \item \textbf{External routines/libraries:} jaxlib, jax, jinja2, matplotlib, numpy, scipy
    \item \textbf{Nature of problem:} The program addresses the need for high-performance computational tools in high-energy physics partial wave analysis (PWA). It introduces the Newton-conjugate gradient method for optimization, enhancing the accuracy and stability of fits.
    \item \textbf{Solution method:} PWACG employs code generation and automatic differentiation techniques to automate the creation of PWA code. It leverages the computational capabilities of JAX for efficient execution and supports advanced features such as multi-GPU computation.
    \item \textbf{GitHub repository:} \url{https://github.com/caihao/PWACG}
\end{itemize}

\section{Introduction}
In high-energy physics experiments, Partial Wave Analysis (PWA) serves as a vital data analysis method for extracting valuable information from measurements~\cite{peters2006primer}. This technique involves analyzing the kinematic variables of final-state particles, particularly in multi-body processes, and integrating this information into the partial wave amplitudes for fitting~\cite{klempt2017partial}. It enables the determination of the presence of intermediate resonant states, along with their mass, width, parity, couplings, and relative phase angles between various amplitudes when considering interference. As a result, a large number of parameters conventionally need to be determined in a PWA~\cite{Krinner2018Ambiguities}, leading to a maximization problem in a high-dimensional parameter space. With the increasing scale of high-energy physics experimental data and the growing complexity of PWA models~\cite{yuan2019besiii}, PWA software faces long-term challenges in data processing capability and speed, thereby attracting sustained interest in its development.

This challenge has been partially resolved by utilizing the continuous advancements in computer technology. Early versions of PWA programs were predominantly developed in C or Fortran language, focusing on CPU-based calculations, exemplified by the PWA2000~\cite{cummings2003object}. Constrained by the computing technology available at the time, these programs could only tackle decay problems involving small data sizes and relatively simple physical models~\cite{aubert2005amplitude}. In recent years, PWA programs utilizing C++ or Python as coding languages and based on Graphics Processing Units (GPUs) have emerged, such as GooFit~\cite{Andreassen_2014} and GPUPWA~\cite{Berger2010Partial}. 
GPUs offer a high degree of parallel processing capability, which, in comparison to traditional CPU programs, can significantly enhance computational speed and efficiency~\cite{Park2009An,Owens2008GPU}, aligning well with the computational requirements of PWA. For instance, the GPUPWA package harnesses GPU parallel computing to expedite the calculation of likelihood functions and their derivatives, achieving a computational speed more than two orders of magnitude faster than traditional FORTRAN code~\cite{Berger2010Partial}. However, with the rapid progress in particle accelerator and detector technologies, the volume of data generated by high-energy physics experiments has grown exponentially~\cite{Apostolakis2022Detector}. This surge in data has made it feasible to identify resonance states that were previously undetectable due to limited statistics, necessitating more complex PWA models to describe the data, entailing a significantly larger number of parameters~\cite{FCpwa}. The concurrent increase in the volume of experimental data and the number of parameters has placed greater demands on the performance of PWA software. Recent advancements also include the development of frameworks such as AmpliTF~\cite{amplitf_github} and ComPWA~\cite{compwa_docs}, which leverage TensorFlow and JAX, respectively, to enhance the efficiency and flexibility of PWA tasks, paving the way for more robust fitting and modeling tools.

Based on the experience in the field of artificial intelligence, it has been observed that the Newton conjugate gradient method~\cite{Yang2009Newton-conjugate-gradient,Sherali1990Conjugate} can accelerate optimization by significantly reducing both computational load and memory usage~\cite{song2022modeling}, in comparison to the traditional standard Newton method. Moreover, while Quasi-Newton methods~\cite{Nocedal2021Quasi-Newton} approximate the Hessian matrix using gradient information, the Newton conjugate gradient method benefits from exact second-order information, leading to higher efficiency in locating global optima.
 Additionally, it has been noted that in the conjugate gradient calculation, automatic differentiation~\cite{baydin2018automatic} is a valuable technique. Unlike the finite difference method~\cite{Mazumder2016The}, automatic differentiation can provide more precise results by minimizing rounding errors in floating-point operations, which is crucial in optimization when determining the descent direction~\cite{Ruder2016An}. However, mainstream PWA software still relies on the Minuit toolkit, including its FORTRAN version Minuit, C++ version Minuit2~\cite{Hatlo2004Developments}, and Python version iMinuit~\cite{iminuit_github}. The Minuit series primarily utilizes quasi-Newton methods for optimization~\cite{James1984MINUIT-a}, such as the BFGS~\cite{Badem2018A} and DFP~\cite{Avriel1976} methods. Even with the development of PWA software like TFPWA~\cite{tfpwa_github}and zfit~\cite{eschle2019zfit}, which implements automatic differentiation, applying the Newton conjugate gradient method to practical PWA tasks remains highly challenging. To enhance PWA efficiency, a well-optimized package that combines the Newton conjugate gradient method and automatic differentiation is expected to significantly improve performance.

In the subsequent section, we will present the Partial Wave Analysis Code Generator (PWACG), which is developed to produce high-performance partial wave analysis programs. The code generated by PWACG is exclusively constructed using the high-performance APIs offered by JAX. It facilitates fitting through the utilization of the Newton conjugate gradient method and multi-GPU collaborative computing to manage extensive data sets. Furthermore, we have conducted comprehensive tests encompassing various data volumes and fitting methods.

\section{Partial Wave Analysis Code Generator}

\subsection{Architecture}
To accurately derive the derivatives of the likelihood function, we turn to popular automatic differentiation tools currently in use. JAX~\cite{jax2018github}, developed by Google, stands out as a high-performance machine learning library. It efficiently compiles and runs Python code using the XLA compiler, making it particularly well-suited for large-scale numerical computations. JAX takes full advantage of modern hardware, such as GPUs and TPUs, to accelerate computations, which is crucial for handling the complex calculations involved in partial wave analysis. Additionally, JAX supports ROCm and Apple's deep learning chips, further broadening its applicability across different hardware platforms. Unlike PyTorch and TensorFlow, which are more focused on neural network computations, JAX can be seen as a GPU/TPU-accelerated version of NumPy. It offers a more fundamental approach in code implementation and superior automatic differentiation capabilities for mathematical functions, making it a natural choice for developing partial wave analysis software.

Throughout the partial wave analysis process, researchers need to continuously try adding or removing intermediate resonances to find the best partial wave analysis model that matches the data. Conventionally, we can introduce control parameters and use branches specific to the control parameters to determine whether to calculate a particular intermediate resonance. When using numerical differentiation to calculate derivatives, we only need to avoid calculating the derivatives of the control parameters.
Nonetheless, to attain high computational performance, our implementation necessarily relies on JAX's just-in-time (JIT) compilation mechanism. JIT compilation works by tracing the execution of a Python function and transforming it into a static computation graph, which is then optimized and executed by the XLA (Accelerated Linear Algebra) compiler. A critical limitation of this static-graph-based approach is its inability to support data-dependent control flow, particularly when expressed using standard Python conditional statements. This is because such constructs introduce dynamic structural changes to the computation graph at runtime. As a result, this constraint fundamentally conflicts with XLA’s requirement for static computation graphs, posing a significant engineering challenge in designing flexible and efficient algorithms.
To build an efficient partial wave analysis software, one feasible approach is to generate the corresponding partial wave analysis program based on the selected combination of resonances using code generation. Jinja2 is a popular Python template engine that provides powerful programming features such as variables, filters, and tags, making it highly suitable for handling complex logic and large templates. In the PWACG framework, we use jinja2 to combine various intermediate resonance templates into a complete partial wave analysis model based on the configuration file provided by the user.

\begin{figure}[htbp]
  \centering
  \includegraphics[width=0.6\textwidth]{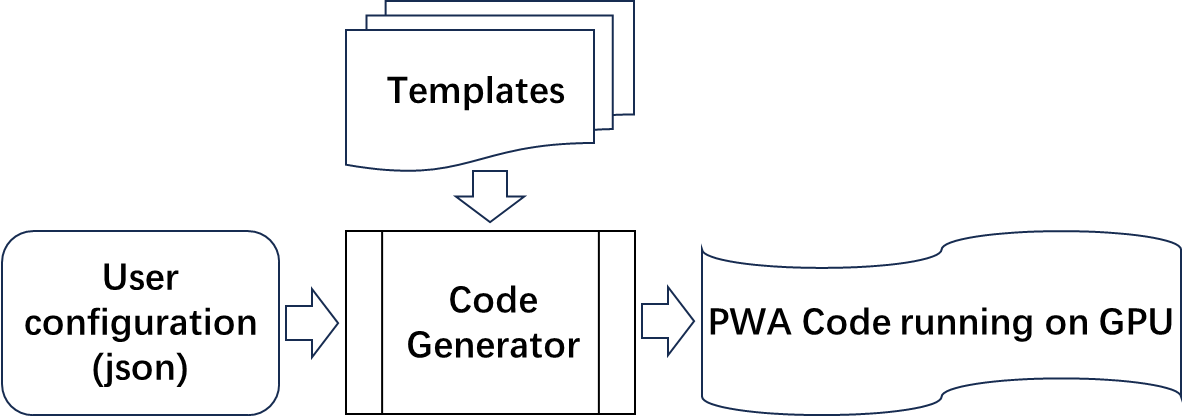}
  \caption{Workflow diagram illustrating the process for generating Partial Wave Analysis (PWA) code optimized for GPU execution. }
  \label{fig-template}
\end{figure}

The PWACG software allows users to flexibly customize all the details of the partial wave analysis program while ensuring the generation of an efficiently executable partial wave analysis program. PWACG consists of the following three components, Figures~\ref{fig-template}:

\begin{enumerate}
    \item User Configuration File (JSON format): Through this configuration file, users specify various parameters and settings for the partial wave analysis program, including the combination of resonances, parameter initialization, constraint conditions, multi-GPU configuration parameters, and more. With the configuration file, users can flexibly obtain the desired partial wave analysis program without directly writing code.
    \item  Template Files: These files contain code templates that define the basic structure of each program module, such as likelihood function computation, data reading and preprocessing, multi-threading and multi-GPU scheduling, and more. The template files serve as the backbone of the entire code generation framework, providing a solid foundation for code generation.
    \item Code Generation Engine: This engine is responsible for combining the user configuration file and template files to generate the final partial wave analysis program. It selects the appropriate templates based on the user's configuration and populates them with the corresponding parameters and settings, thereby generating code that can run directly on GPUs.
\end{enumerate}

\subsection{Implementation and Performance Optimization of Computational Templates}
PWACG's computation templates are mainly divided into two categories: one for calculating the amplitude of a single intermediate resonance, and the other for assembling these individual templates into a complete partial wave analysis computation model. Currently, PWACG is based on the covariant tensor formalism to construct the partial wave analysis model, and its implementation and optimization approach can also be applied to the helicity formalism.

The differential cross-section for particle decay can be represented as:
\begin{equation}
\omega(\xi_i,\alpha) = \frac{d \sigma}{d \Phi} = \frac{1}{2} \sum_{\mu=1}^2 A^\mu A^{* \mu} = \frac{1}{2} \sum_{m, n} \Lambda_m \Lambda_n^* \sum_{\mu=1}^2 U_m^\mu U_n^{* \mu}\tag{1}
\end{equation}
where $\xi_i$ represents the $i$-th event, $\alpha$ is the undetermined parameter in the partial wave analysis model, the total amplitude $A$ is obtained by summing the partial wave amplitudes of all intermediate resonances, and the partial wave amplitude of an intermediate resonance is obtained by multiplying the coupling strength $\Lambda_i$ by the projection of the partial wave amplitude in a specific direction, $U_i^\mu$. The likelihood function for partial wave analysis is:
\begin{equation}
L = \prod_{i=1}^{N_{Data}} \frac{\omega(\xi_i, \alpha)}{\int d \xi \omega(\xi, \alpha) \epsilon(\xi)} \tag{2}
\end{equation}
Here, the product is taken over all $N_{\text{Data}}$ events in the data sample, $\epsilon(\xi)$ is the detector efficiency, and the integral is obtained by summing over $N_{MC}$ simulated phase-space Monte Carlo events. The log-likelihood function is expanded as:
\begin{equation}
\ln L = \sum_{i=1}^{N_{Data}} \ln \left(\sum_{\mu=1}^2 A^\mu(\xi_i) A^{* \mu}(\xi_i)\right) - N_{Data}\ln \left(\frac{1}{N_{MC}}\sum_{j=1}^{N_{MC}} \sum_{\mu=1}^2 A^\mu(\xi_j) A^{* \mu}(\xi_j)\right) \tag{3}
\end{equation}
From equation $(1)$, we can see that the total amplitude $A$ is mainly calculated according to the Einstein summation convention, and JAX provides the $\texttt{einsum}$ operation to support the implementation of the Einstein summation convention. In equation $(1)$, $U_i^\mu$ does not contain any undetermined parameters, meaning that it does not need to be repeatedly calculated during the optimization process of the entire partial wave analysis model, and can therefore be pre-computed and stored on the GPU. To fully utilize the tensor computation capabilities of the GPU, we use JAX's $\texttt{vmap}$ operation to completely tensorize equation $(3)$. All of the above computations are fixed through templates, and to obtain the complete partial wave analysis computation program, one only needs to perform template replacement and assembly.

After generating the partial wave analysis code, JAX can further optimize the computation process by utilizing the XLA (Accelerated Linear Algebra) compiler, such as automatic parallelization and memory access optimization, thus fully unleashing the computational power of the GPU. XLA can optimize these operations by transforming vectorized function computations into linear algebra operations, such as merging multiple operations to reduce memory access overhead, or rearranging the order of operations to improve computational efficiency. We have also made some other optimizations to improve computational efficiency, notably overloading the construction and storage method for complex number arrays. Specifically, we store the real and imaginary parts of complex numbers separately in a two-dimensional array, and then perform all related computations using JAX's tensor computation tools. This approach significantly improves the efficiency of automatic differentiation compared to directly using complex numbers for computation.

\subsection{Multi-GPU Support}
Nowadays, the data volume that needs to be processed in high-energy physics experiments for partial wave analysis is rapidly increasing. On the other hand, with the introduction of the Newton conjugate gradient method and Hessian-vector products (HVP), the limited memory on a single GPU often cannot meet the computational demands of partial wave analysis software. Researchers have thus started to consider utilizing multiple GPUs for parallel computation. If only first-order derivatives are required in the computation process, parallel computation across multiple GPUs is relatively straightforward to implement. However, in PWACG, we need to support HVP computation across multiple GPUs, which is a more complex problem.

Suppose we have $M$ GPUs. We divide the experimental data and Monte Carlo simulation data into $M$ parts and store them separately on the GPUs. We denote the experimental data volume on each GPU as $N_{Data,m}$, and the Monte Carlo simulation data volume on each GPU as $N_{MC,m}$. Then, equation (3) can be written as:
\begin{equation}
\ln L = \sum_{m=1}^MF_m - N_{\text{Data}} \ln \left( \frac{1}{N_{\text{MC}}} \sum_{m=1}^MG_m \right)\tag{4}
\end{equation}
where
\begin{equation}
F_m=\sum_{i=1}^{N_{Data,m}} \ln \left(\sum_{\mu=1}^2 A^\mu(\xi_i) A^{* \mu}(\xi_i)\right),~G_m= \sum_{j=1}^{N_{MC,m}} \sum_{\mu=1}^2 A^\mu(\xi_j) A^{* \mu}(\xi_j)\tag{5}
\end{equation}
Taking the first-order derivative of equation (4), we have:
\begin{equation}
\frac{\partial\ln L}{\partial\alpha_i}=\sum_{m=1}^M\frac{\partial F_m}{\partial\alpha_i}-N_{Data}\frac{1}{\sum_{m=1}^MG_m}\sum_{m=1}^M\frac{\partial G_m}{\partial\alpha_i}\tag{6}
\end{equation}
Continuing to obtain the second-order derivative:
\begin{equation}
\frac{\partial^2\ln L}{\partial\alpha_i\partial\alpha_j}=\sum_{m=1}^M\frac{\partial^2F_m}{\partial\alpha_i\partial\alpha_j}-\frac{N_{Data}}{\sum_{m=1}^MG_m}\frac{\partial^2 G_m}{\partial\alpha_i\partial\alpha_j}+\frac{N_{Data}}{\left(\sum_{m=1}^MG_m\right)^2}\sum_{m,n=1}^M\frac{\partial G_m}{\partial\alpha_i}\frac{\partial G_n}{\partial\alpha_j}\tag{7}
\end{equation}
From equation (7), we can see that to obtain the HVP of $\ln L$, we only need to compute $G_m$, the first-order derivative of $G_m$, and the HVP of $F_m$ and $G_m$ on each GPU separately. Based on the above computation formulas, we have implemented parallel computation across multiple GPUs on a single compute node using Python's multi-threading capabilities. In the future, if needed, we can implement parallel computation across nodes to support partial wave analysis on an even larger scale.

\section{Performance Test}
PWACG generates high-performance PWA code that can rapidly and accurately compute the likelihood, first-order derivative, and HVP of any PWA model. This makes it suitable for various optimization methods. To comprehensively evaluate the performance of PWACG, we constructed a PWA model to describe the decay process $\psi(3686) \rightarrow \phi K^{+} K^{-}$~\cite{ZouBS}. This model encompasses intermediate processes such as $\psi(3686) \to \phi f$, $f \to K^+ K^-$, where $f$ represents various scalar or tensor resonances including $f_0(980)$~\cite{flatte1,flatte2}, $f_2(1270)$~\cite{Shchegelsky:2006et}, $f_2^{\prime}(1525)$~\cite{Longacre:1986fh}, $f_0(1710)$~\cite{Belle:2013eck}, and $f_2(2150)$~\cite{WA102:2000lao}. The fraction, mass, and width of these resonances are detailed in Table 1, while the phase angles and couplings of individual partial wave amplitudes are not included here for brevity. A Monte Carlo (MC) sample comprising 10,000 events based on this physical model was generated. The corresponding Dalitz plot and $K^{+}K^{-}$ invariant mass spectrum are depicted in Fig. 1 and Fig. 2, respectively. This model encompasses both broad and narrow resonances, with relatively strong interference between the resonances, presenting a challenging scenario for PWA. It is important to note that this MC sample set is utilized for meticulously testing the performance of the PWA program, rather than providing a complete representation of the actual three-body decay $\psi(2S) \rightarrow \phi K^{+} K^{-}$.

\begin{table}[htbp]
\centering
\caption{Resonances and their parameters in the PWA model.}
\begin{tabular}{ccccc}
\toprule
$R_0$ & Name & $F_i$(\%) & Mass (GeV) & Width (GeV) \\
\midrule
1 & $f_{0}(980)$ & 42.89 & 1.022 & 0.209 \\
2 & $f_{2}(2550)$ & 30.95 & 2.545 & 0.303 \\
3 & $f^{'}_{2}(1525)$ & 30.27 & 1.522 & 0.091 \\
4 & $f_{0}(1710)$ & 8.97 & 1.672 & 0.169 \\
5 & $f_{2}(1270)$ & 3.98 & 1.302 & 0.189 \\
6 & $f_{2}(2150)$ & 2.52 & 2.150 & 0.143 \\
\bottomrule
\end{tabular}
\label{tab-1}
\end{table}

\begin{figure}[htbp]
  \centering
  \includegraphics[width=0.6\textwidth]{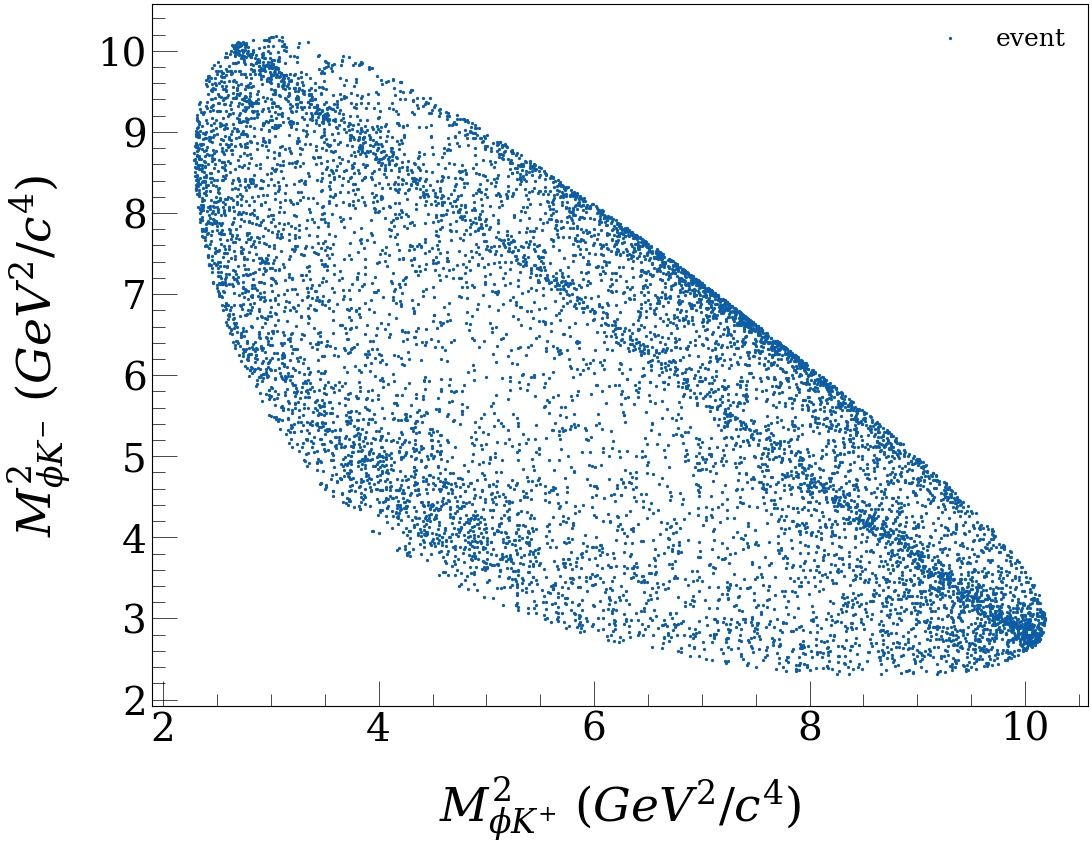}
  \caption{The Dalitz plot from the MC sample.}
  \label{fig-dalitz}
\end{figure}

\begin{figure}[htbp]
  \centering
  \includegraphics[width=0.6\textwidth]{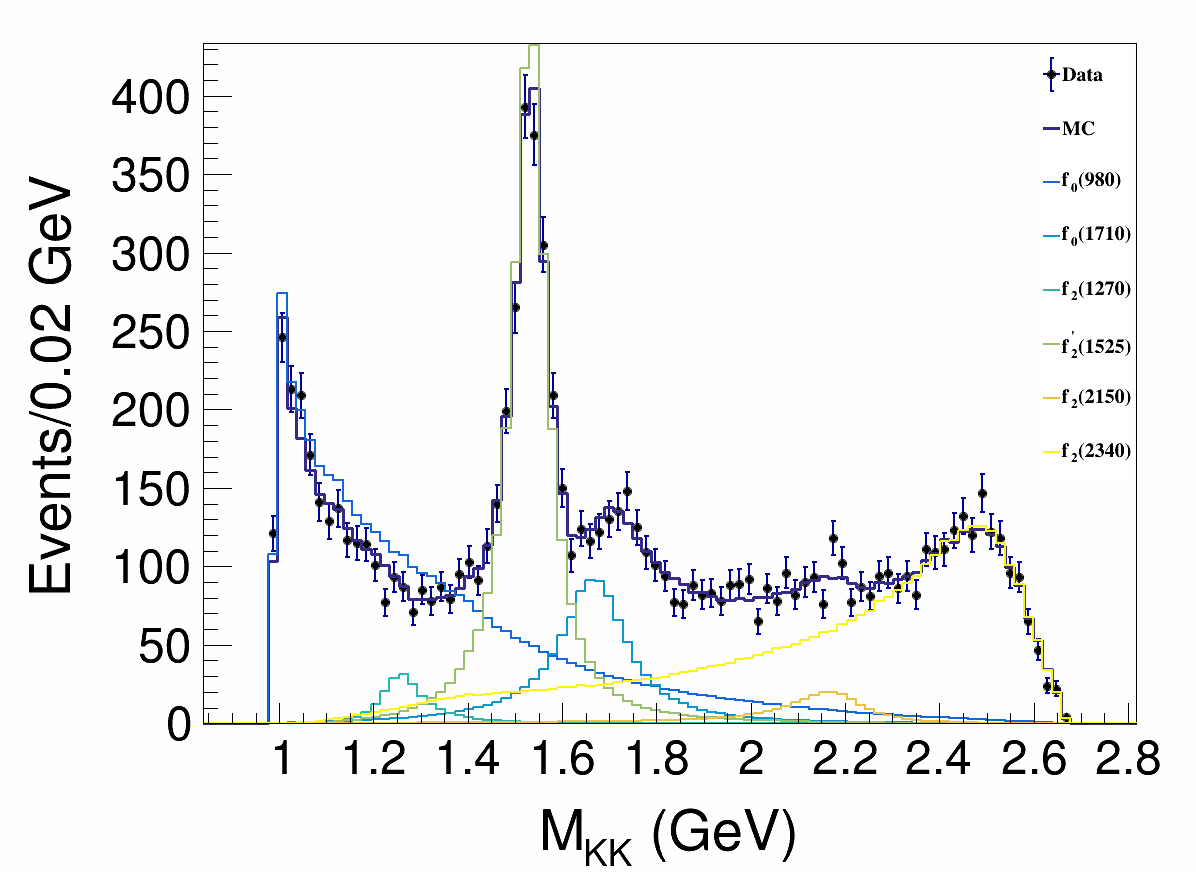}
  \caption{The $K^+ K^-$ invariant mass spectrum from the MC sample.}
  \label{fig-weight}
\end{figure}

In the conducted tests, we utilized the GPU cluster at the Super Computing Center of Wuhan University, with each compute node being equipped with 4 Nvidia Tesla V100 GPUs, each boasting 16GB of memory. It's worth noting that the Nvidia Tesla V100 GPU, released in 2017, implies that running the codes with subsequently released GPUs will significantly enhance the performance of PWACG. The primary focus of our testing involved the comparison of the following three cases:
\begin{itemize}
\item iMinuit+FDM: Calculate derivatives using finite difference method(FDM), and use iMinuit for optimization. 
\item iMinuit+AutoDiff: Calculate first-order derivatives using automatic differentiation, and use iMinuit for optimization.
\item NTCG+AutoDiff: Calculate first-order derivatives and HVP using automatic differentiation, and use $\texttt{scipy.optimize(method="Newton-CG")}$ for optimization.
\end{itemize}



\begin{figure}[htbp]
  \centering
  \begin{subfigure}[b]{0.48\textwidth}
    \centering
    \includegraphics[width=\textwidth]{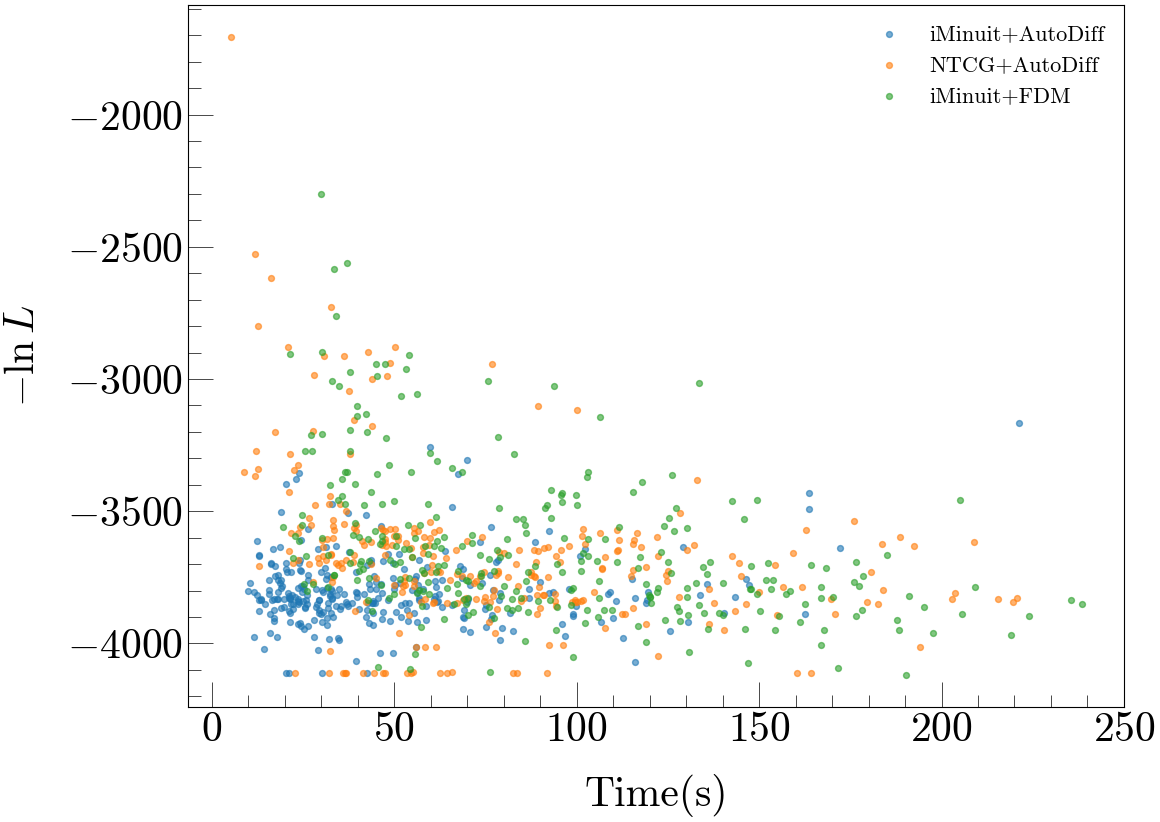}
    \caption{Scatter plot of $-\ln L$ versus time.}
    \label{fig-bm1}
  \end{subfigure}
  \hfill
  \begin{subfigure}[b]{0.48\textwidth}
    \centering
    \includegraphics[width=\textwidth]{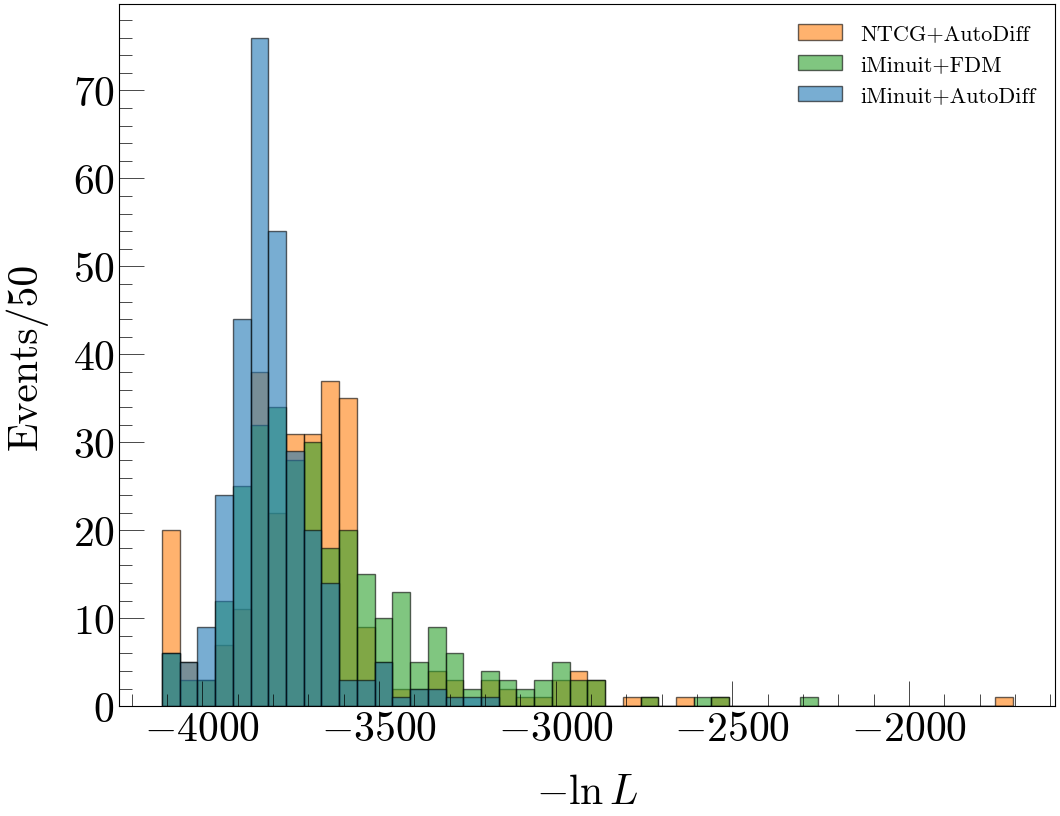}
    \caption{$-\ln L$ distributions of 300 fits.}
    \label{fig-bm2}
  \end{subfigure}
  \caption{The results of 300 fits for three different optimization methods. The convergence precision of FCN for method iMinuit+FDM was set to $10^{-2}$ , while the methods iMinuit+AutoDiff and NTCG+AutoDiff was set to $10^{-5}$.}
  \label{fig4}
\end{figure}

We randomly generated 300 sets of parameters and used them as the fitting starting points with the three methods. 
A fundamental criterion for a valid performance comparison of these optimization methods is to evaluate them under an equivalent convergence precision. In our example PWA model, our empirical tests reveal significant differences in the maximum achievable precision for each approach: 
\begin{itemize}
    \item iMinuit+FDM: The convergence of the function to be minimized (FCN) is limited to a precision of approximately $10^{-2}$. 
    Setting a smaller value leads to convergence difficulties and increases the computation time, without yielding any improvement in precision. In iMinuit, the parameter $tol$ can be adjusted to affect the max estimated distance to minimum (EDM), thereby controlling the convergence precision~\cite{iminuit_web}.
    \item iMinuit+AutoDiff: Since AutoDiff provides more accurate derivatives, this method can achieve an FCN convergence precision up to $10^{-5}$.
    \item NTCG+AutoDiff: In the NTCG fitting method provided by \texttt{scipy.optimize}, we influence the convergence precision of FCN by controlling the convergence precision of fitting parameters $xtol$~\cite{SciPyOptimize}. The default value of $xtol$ is $10^{-8}$, and we can relax $xtol$ restriction to provide a more reasonable comparison condition with the two methods above.
\end{itemize}
%
In Figure~\ref{fig4}, we compare the fitting capabilities of these three methods. Here we set the $xtol$ parameter of NTCG+AutoDiff to $10^{-5}$ to achieve an FCN convergence precision close to that of iMinuit+AutoDiff.
The distributions of the resulting $-\ln L$ values are shown in Fig.~\ref{fig-bm1}. The global optimum value of $-\ln L$ is $-4110$, with results below $-4109$ considered the global optimum for statistical purposes. Among these methods, NTCG+AutoDiff has the highest success rate in finding the global optimum at 6.67\%, followed by iMinuit+AutoDiff at 1.33\%, while iMinuit+FDM has a rate of only 0.33\%, as shown in Fig.~\ref{fig-bm2}.

\begin{figure}[htbp]
  \centering
  \begin{subfigure}[b]{0.48\textwidth}
    \centering
    \includegraphics[width=\textwidth]{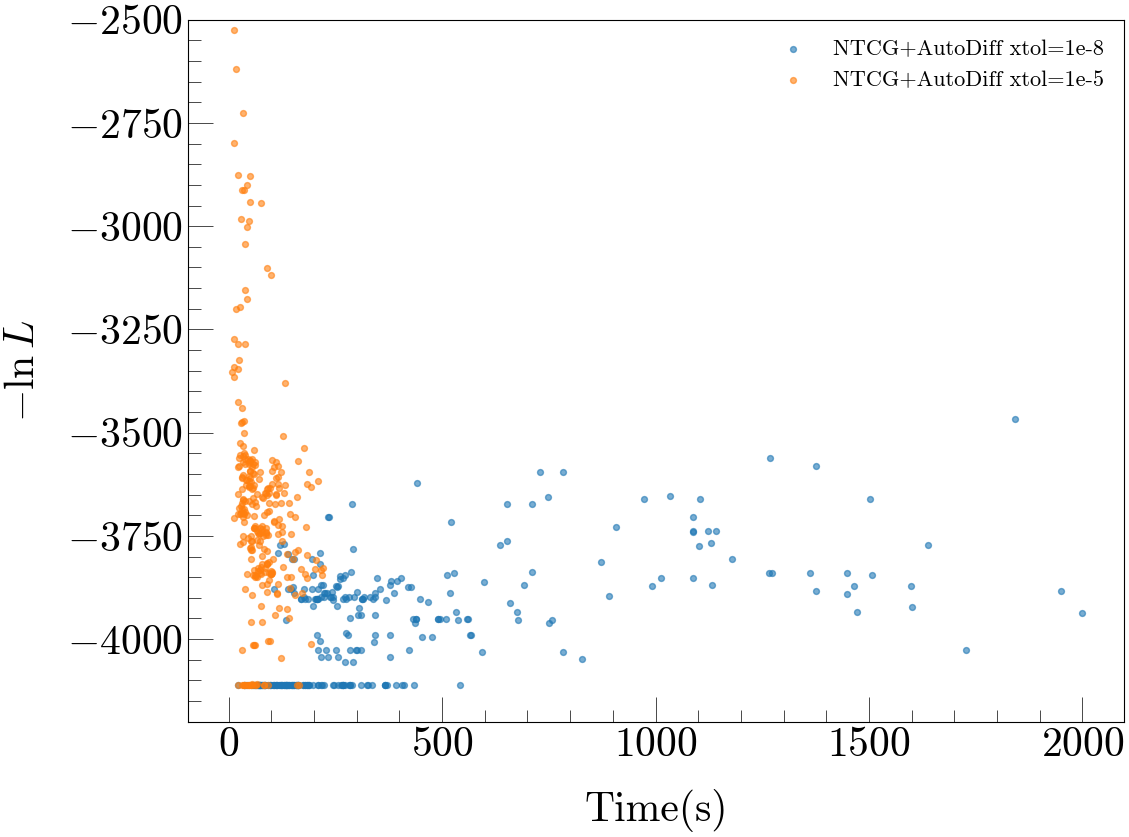}
    \caption{Scatter plot of $-\ln L$ versus time.}
    \label{fig-bm1-NTCG}
  \end{subfigure}
  \hfill
  \begin{subfigure}[b]{0.48\textwidth}
    \centering
    \includegraphics[width=\textwidth]{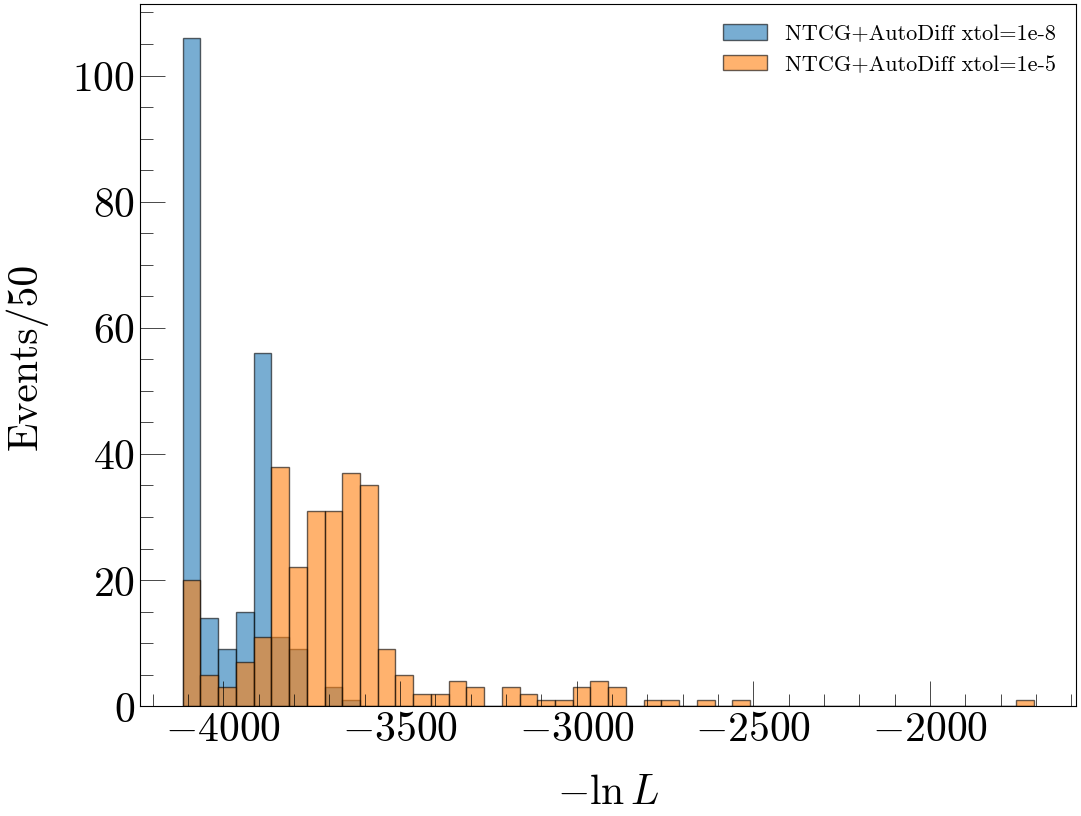}
    \caption{$-\ln L$  distributions of 300 fits.}
    \label{fig-bm2-NTCG}
  \end{subfigure}
  \caption{Performance comparison of NTCG+AutoDiff at $xtol=10^{-8}$ and $xtol=10^{-5}$}
\end{figure}

In actural PWA probelm, NTCG+AutoDiff allows us to set a much lower $xtol$, below $10^{-8}$. This setting effectively helps in avoiding local minimal, ultimately achieving a success rate of 35\%, as shown in Fig.~\ref{fig-bm1-NTCG} and Fig.~\ref{fig-bm2-NTCG} . Additionally, for NTCG+AutoDiff, an ``early stopping~\cite{dossa2021empirical}'' strategy similar to deep learning optimization, where the fitting process is terminated if it exceeds 500 seconds, can be adopted, minimally impacting the success rate of finding the global optimum while significantly saving time.

\begin{figure}[htbp]
  \centering
  \includegraphics[width=0.6\textwidth]{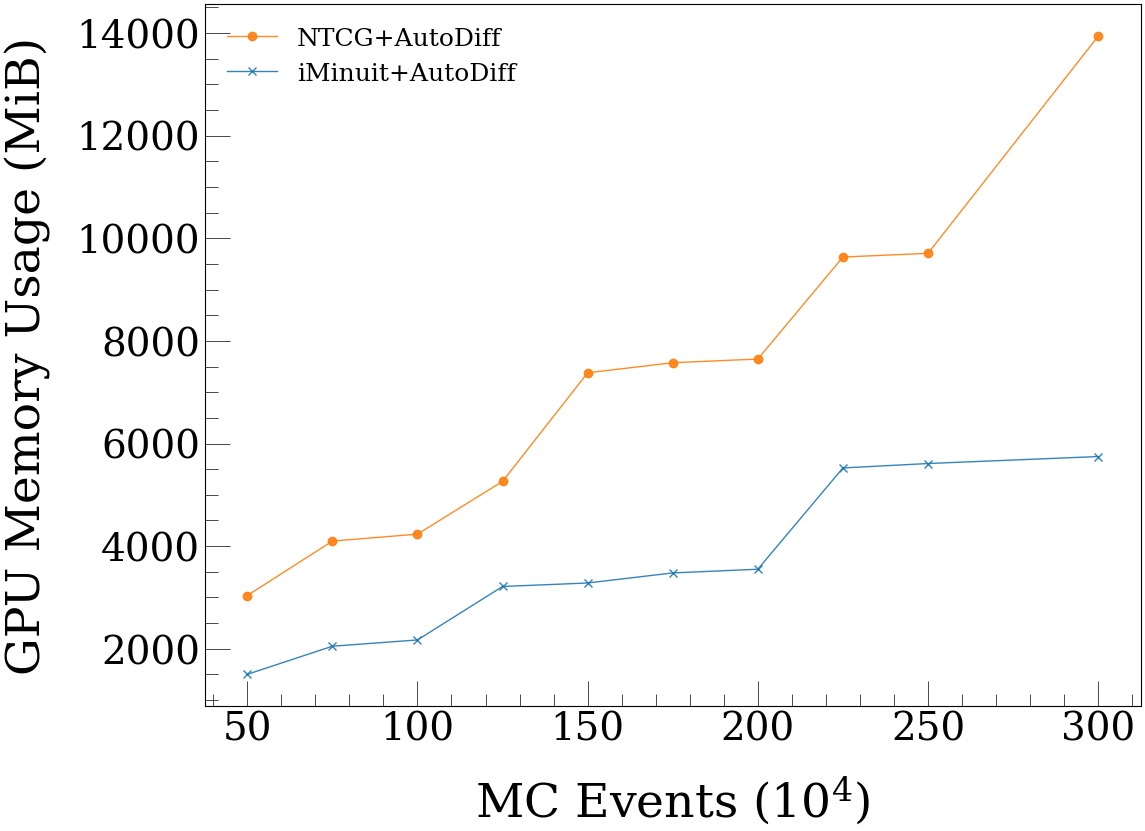}
  \caption{GPU memory consumption for NTCG and Migrad optimization methods with various sizes of MC sample.}
  \label{fig-mem}
\end{figure}

While NTCG+AutoDiff excels in finding the global optimum, its requirement for computing the HVP results in significantly higher memory usage compared to methods that only compute the first-order derivative, imposing limitations on its use. To facilitate comparison, we evaluated the memory usage under different data scales by varying the number of MC phase-space sample events used for integration, while keeping the model unchanged. As depicted in Fig.~\ref{fig-mem}, when the data volume is relatively small, the memory usage of NTCG+AutoDiff is approximately twice that of the iMinuit+AutoDiff method. Moreover, as the number of integration sample events increases, the memory usage of NTCG+AutoDiff grows at a faster rate. When the number of integration sample events reaches 3,000,000, the memory usage exceeds the 16GB capacity of the V100 GPU, leading to overflow. However, practical PWA typically does not require more than 1,000,000 integration sample events. Therefore, even for slightly more complex PWA models, NTCG+AutoDiff is sufficient to meet practical needs. In the event that future PWAs necessitate larger data volumes and correspondingly larger-scale integration samples, PWACG can effectively support this through multi-GPU parallelization.

\begin{figure}[htbp]
  \centering
\includegraphics[width=0.6\textwidth]{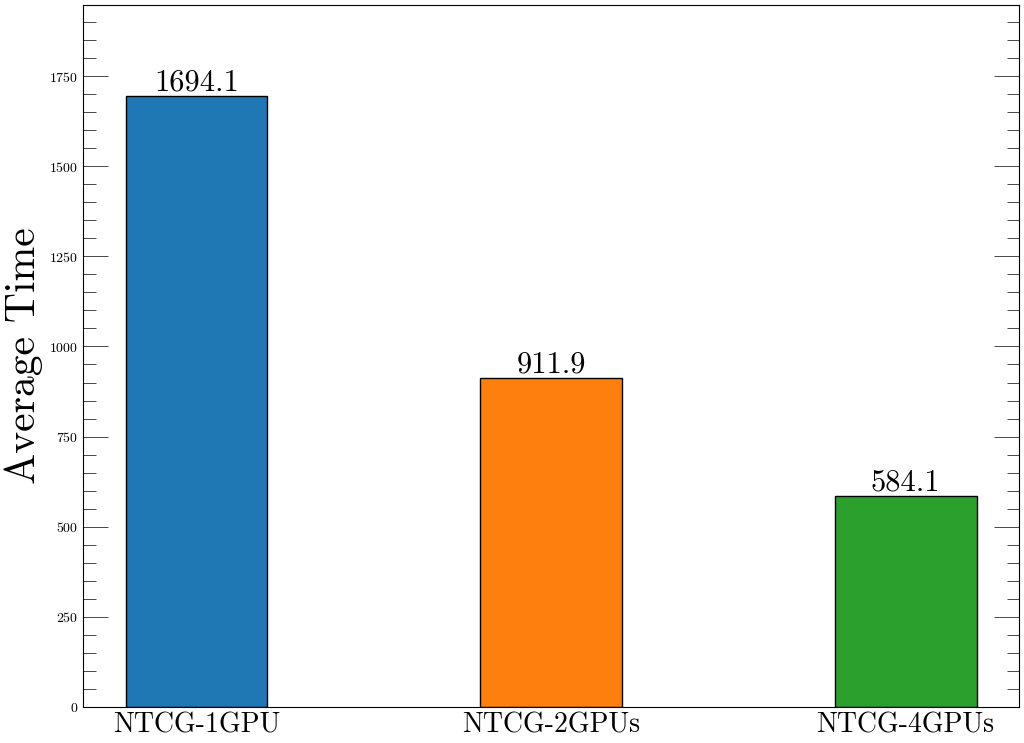}
\caption{Average time consumed for parallel fitting under different GPU configurations.}
\label{fig-ave-bm5}
\end{figure}

Multi-GPU parallel computing can effectively mitigate the limitations of insufficient memory on a single GPU and substantially enhance fitting efficiency. In order to fully showcase the acceleration capabilities of multi-GPU parallel computing, we conducted parallel tests using different numbers of GPUs while maintaining the model and data scale unchanged. It is important to note that, at present, PWACG exclusively supports multi-GPU computation on a single node. However, in the future, support for multi-GPU parallelism across nodes could be implemented through MPI. 
Fig.~\ref{fig-ave-bm5} illustrates the average time consumed in different numbers of GPUs under the condition of 1,200,000 integration instances. The results demonstrate that PWACG has undergone significant optimizations for multi-GPU parallel computing, with multiple GPUs delivering substantial performance improvements.

\section{Summary}

PWA plays a crucial role in the data analysis of high-energy physics experiments, serving to determine the physical parameters of intermediate resonant states. However, traditional PWA software encounters computational inefficiency challenges when handling large-scale data and complex models. PWACG addresses this issue by harnessing the JAX automatic differentiation library and the jinja2 template engine to automatically generate high-performance PWA code based on user configurations. The resulting code is exclusively built on the high-performance APIs of JAX, supports optimization through the Newton conjugate gradient method, and fully capitalizes on GPU parallel computing capabilities. 

The paper provides an in-depth overview of the software architecture of PWACG, the implementation and optimization of computational templates, and the support for multi-GPU parallelism. It evaluates and compares the performance of PWACG-generated code, in combination with various optimization algorithms, from multiple perspectives such as likelihood function convergence, memory consumption, and multi-GPU acceleration.

Thanks to substantial optimizations in computational performance and memory usage, the PWACG-generated code supports optimization using the Newton conjugate gradient method for large data sets. In comparison to traditional PWA software utilizing other optimization methods, PWACG significantly outperforms in efficiency in finding global optima, a benefit that is expected to further amplify with advancements in computing hardware. The design philosophy, key technological implementations, and performance of the PWACG automated PWA code generation framework introduce a new approach to enhancing the computational efficiency of PWA.

\section*{Acknowledgments}
The authors would like to express their gratitude to the Supercomputing Center of Wuhan University for providing the supercomputing resources that have contributed to the numerical calculations reported in this paper. 

\noindent This work was supported by the National Science Foundation of China under Grant Nos. 11735010, U1932108, U2032102, and 12061131006. Their financial support was instrumental in the realization of our research goals.

\section*{Data and Software Availability}
The source code for the custom software and scripts used in this study has been made publicly available to ensure transparency and reproducibility of our research. The code is hosted on GitHub at \url{https://github.com/caihao/PWACG}.

\bibliographystyle{unsrt}

\bibliography{sample}

\end{document}